\documentclass[]{aa}
\usepackage{graphicx}
\usepackage{color}
\usepackage{txfonts}

\begin{document}

\title{Quasi-simultaneous five-frequency VLBA observations of PKS 0528+134}
\author{  H. -B. Cai    \inst{1,2}
       \and Z. -Q. Shen     \inst{1,3}
       \and X. Chen  \inst{1,2}
       \and L. -L. Shang \inst{1,2}}%
\offprints{H.-B. Cai, \email{hbcai@shao.ac.cn}}
\institute{Shanghai Astronomical Observatory, Chinese Academy of
Sciences, Shanghai 200030, China \and Graduate School of Chinese
Academy of Sciences, Beijing 100039, China \and Joint Institute
for Galaxy and Cosmology, SHAO and USTC, China}

\date{Received 27 February 2006 / Accepted 28 July 2006}

\abstract {We present results of Very Long Baseline Array (VLBA)
observations of PKS 0528+134 at five frequencies (2.3, 5.0, 8.4,
15.4, and 22.2 GHz). These quasi-simultaneous data enable us to
study the spectral distribution of Very Long Baseline Interferometer
(VLBI) components for the first time in this highly variable source,
from which the central compact core is identified. Our observations
indicate that there are two bendings for the jet motion at parsec
scale. We provide an approximate spatial fit to the curved jet
trajectory using the Steffen et al. (\cite{Steffen95}) helical
model. We further investigate the proper motions of three jet
components, which all show superluminal motion. At high frequencies
(15.4 and 22.2 GHz) we detected a new component, which is estimated
to be related to a radio burst peaking at about 2000.

} \maketitle\markboth{Quasi-simultaneous
five-frequency VLBA observations of PKS 0528+134} {H. -B. Cai et
al.}

\maketitle

\section{Introduction}

As one of the brightest active galactic nuclei, \object{PKS
0528+134} has been intensively studied at the radio, X-ray, and
gamma-ray bands. It is one of the furthest blazars detected by EGRET
above 100 Mev (Mukherjee et al. \cite{Mukherjee97}) with red-shift
z=2.07 (Hunter et al. \cite{Hunter}). There are relatively less
optical observations compared with the observations at other bands
because of its location close to the Galactic plane (galactic
latitude b=$-11\degr0'43''$) and thus high Galactic extinction.
Whiting et al. (\cite{Whiting03}) obtained $B=17.04\pm0.23$,
$V=16.74\pm0.15$ at 2001.66 after the extinction corrections. But
these are different from the previous values B=20.0 (Condon, Hicks
\& Jauncey \cite{Condon77}) and V=19.5 (Wall \& Peacock
\cite{Wall85}), implying its optical variability. People have
observed its variability on a timescale as short as a few days at
gamma-ray (Mukherjee et al. \cite{Mukherjee96}), as short as two
days in the optical band (Ghosh et al. \cite{Ghosh00}), and from
several months to a few years in the radio band (e.g., Pohl et al.
\cite{Pohl95}; Peng et al. \cite{Peng01}). There is a delay for the
radio burst from high frequencies to low frequencies. Peng et al.
(\cite{Peng01}) showed in their light curves that the delay for the
1996 burst is about 0.7 yr between 10.7 and 5.0 GHz. Mukherjee et
al. (\cite{Mukherjee96}) showed that the delay between the gamma-ray
flare and the radio burst is a few months.

Very Long Array (VLA) observations detected a diffuse emission 2$''$
in extent and centered 1$''$ north of the core with $<$ 0.2\% and
1.2\% of the peak brightness at 4.9 and 1.5 GHz, respectively
(Perley \cite{Perley82}). \object{PKS 0528+134} is frequently
observed as a calibrator for VLBI observations (e.g. Shen et al.
\cite{Shen05}). The first VLBI images of \object{PKS 0528+134}
(Charlot \cite{Charlot90}) show a one-sided extension at 2.3 GHz and
a double structure at 8.4 GHz, indicating a core-jet structure at
P.A. (position angle) $\sim-140\degr$. Zhang et al. (\cite{Zhang94})
studied in detail the geometry, radiation mechanism, and physical
environment of the emission region in the source through the X-ray
and VLBI observations, and found a bent jet extending toward the
northeast on a parsec scale. Krichbaum et al. (\cite{Krichbaum95},
hereafter K95) presented VLBI observations at 8, 22, 43 and 86 GHz,
and also revealed a strongly bent one-sided core-jet structure with
at least three superluminal components and two apparently stationary
jet components. Their first 86 GHz VLBI observations of this source
revealed an unresolved core ($\le$50$\mu$as) and a new jet component
(N2) whose ejection was believed to relate to an outburst in the
mm-regime and a preceding gamma-ray flare observed in early 1993.
Britzen et al. (1999, hereafter B99) studied in detail the
morphology, kinetics, and the relations between the morphological
changes in the source structure and the increased activity in the
radio and gamma-ray bands through the eight-year 8 GHz geodetic VLBI
observations composed of 20 epochs. They found seven distinct jet
components that all showed superluminal motions. The jet trajectory
was bent twice, and different components had different paths. But
these radio observations cannot give the spectral distribution of
VLBI components, because the source is strongly variable, while VLBI
observations made at different frequencies were not simultaneous.

In Sect. 2, we present the first quasi-simultaneous VLBI mapping
observations of \object{PKS 0528+134} at five frequencies (2.3, 5.0,
8.4, 15.4, and 22.2 GHz). Then in Sect. 3 we discuss the structure
of the jet, give the spectra of the jet components, identify the
central core component based on its flat spectrum characteristic,
study the proper motions of the jet components, estimate the
ejection epochs of two components (a and n2), and describe the jet
bending by means of the helical model, followed by the discussion
and conclusion in Sects. 4 and 5, respectively.

Throughout this paper, the radio spectral index $\alpha$ is defined
as $S_{\nu}\propto\nu^{-\alpha}$. By assuming $H_0=71$ kms$^{-1}$
 Mpc$^{-1}$, $\Omega_M=0.27$, and
$\Omega_\Lambda=0.73$ (Spergel et al. \cite{Spergel03}), we have a
scale for \object{PKS 0528+134} (z = 2.07) of 1 mas $=8.45 $ pc.

\section{Observations and data reduction}

The observations were performed with the NRAO\footnote{The National
Astronomy Observatory (NRAO) is operated by Associated Universities
Inc., under cooperative agreement with the National Science
Foundation.} Very Long Baseline Array (VLBA) on August 20, 2001
(2001.64) at 2.3/8.4 (dual frequency), 5.0, 15.4, and 22.2 GHz.
Observations at different frequencies were interlaced, so data at
different frequencies have similar (u,v) coverage. At 2.3/8.4 and
5.0 GHz, each scan lasted about 2.6 mins, with total observing time
about 36 mins each. At 15.4 and 22.2 GHz, scan lengths are about 25
and 8 seconds with a total observing time of 60 and 38 mins,
respectively. The data were recorded in 1-bit sampling VLBA format
with a total bandwidth of 64 MHz (eight 8MHz IF channels) at 5.0,
15.4, and 22.2 GHz, and the total bandwidth of 32 MHz (four 8 MHz IF
channels ) at the dual frequency 2.3/8.4 GHz, per circular
polarization at each station.

\begin{figure*}
\resizebox{\hsize}{!}{\rotatebox{0}{\includegraphics{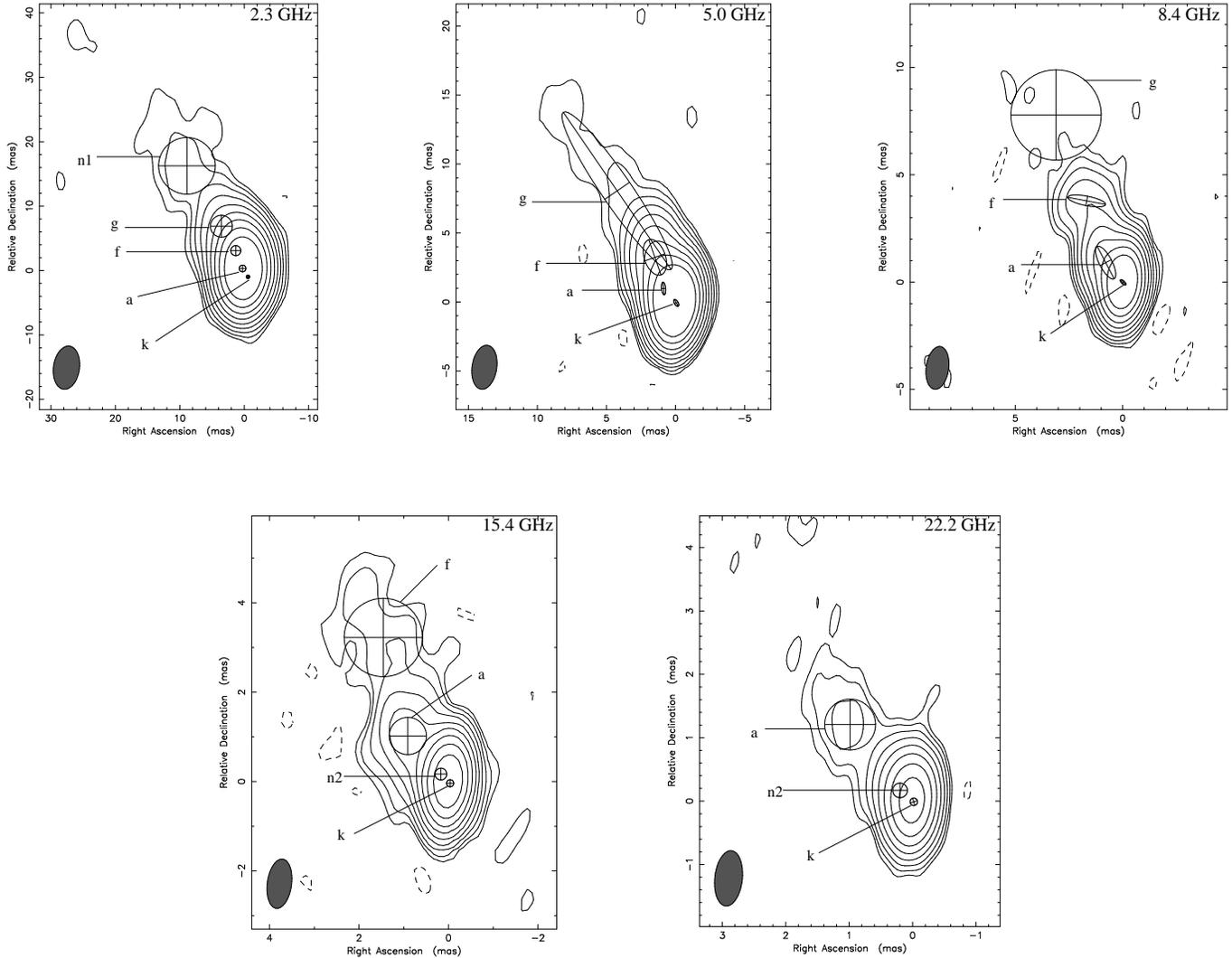}}}
\caption{Naturally weighted maps of \object{PKS 0528+134} at 2.3,
5.0, 8.4, 15.4, and 22.2 GHz. The fitting model components are shown
with the corresponding names. The peak flux densities from low
frequency to high frequency are 1.64, 1.97, 1.8, 1.65, and 1.26
Jy/beam. The beam sizes and orientations from low frequency to high
frequency are 6.77$\times$4.04 mas$^2$ and -7.97$\degr$,
3.22$\times$1.79 mas$^2$ and -8.60$\degr$, 2.01$\times$1.05 mas$^2$
and -7.78$\degr$, 1.11$\times$0.54 mas$^2$ and -7.54$\degr$,
0.88$\times$0.44 mas$^2$ and -6.1$\degr$. The off-source rms noise
levels ($\sigma$) are 0.64, 0.51, 0.82, 1.0, and 2.1 mJy/beam.
Contour levels are drawn at 3$\sigma$ $\times$(-1, 1, 2, 4, 8, 16,
32, 64, 128) at 22.2 GHz, and 3$\sigma$ $\times$(-1, 1, 2, 4, 8, 16,
32, 64, 128, 256) at other frequencies. } \label{fig:1}
\end{figure*}

The data correlation was done with the VLBA correlator in Socorro,
New Mexico, USA. Post-correlation data was exported to the NRAO AIPS
software (Schwab \& Cotton \cite{Schwab83}) for fringe fitting, then
to the Caltech DIFMAP package (Shepherd \cite{Shepherd97}) for
hybrid mapping. A prior visibility amplitude calibration was done
using the antenna gain and the system temperature measured at each
station. During the hybrid mapping, the CLEAN and phase-only
self-calibration were iteratively used during the early processing,
and the amplitude self-calibration was added for the later
processing. The resulting VLBA maps at five frequencies are shown in
Fig.~\ref{fig:1}. The quantitative description of the source
structure is determined by model fitting to the calibrated
visibility data at each frequency, and the results listed in
Table~\ref{tab:1}.

We used the Difwrap program (Lovell \cite{Lovell00}) to estimate the
errors in Gaussian model fitting. The parameter of interest was
stepped in small increments near the best-fit value. At each step
this parameter was kept fixed, while some related parameters were
allowed to relax during the model fitting. In addition, errors
caused by self-calibration during the mapping processes had to be
considered. These are typically 10\% at 2.3, 5.0, 8.4, and 15.4 GHz,
and about 15\% at 22.2 GHz. The errors listed in Table~\ref{tab:1}
were obtained by combining all these error budgets.

The single-dish flux density at the time close to our VLBA
observations is 2.62$\pm0.08$, 2.40$\pm0.08$, and 2.52$\pm0.04$ Jy
at 5.0, 8.4, and 15.4 GHz, respectively, from the University of
Michigan Radio Observatory, 2.20$\pm0.03$ Jy at 2.3 GHz from the
NRAO Green Bank Interferometer, and 2.14 Jy at 22.2 GHz from the
Metsahovi Observatory. The total flux density observed by these
single dishes is 1.07$\pm0.01$, 1.06$\pm0.03$, 1.06$\pm0.04$,
1.16$\pm0.02$, and 1.26$\pm0.08$ times higher than the integrated
flux density in the VLBA maps at 2.3, 5.0, 8.4, 15.4, and 22.2 GHz,
respectively. We note that the relatively high ratio at 22.2 GHz was
also seen at three epochs of VLBA observations of \object{PKS
0528+134} by Ojha et al. (\cite{Ojha04}). These corrections to the
absolute flux densities at different frequencies are applied in
Sect. 3.2, where the spectral indexes of VLBI components are
estimated.

\begin{table*}
\centering \caption{The results of model-fitting. } \label{tab:1}
\begin{minipage}{155mm}
\vspace{2mm}
\begin{tabular}{cccccccc}\hline\hline

Component &Flux(Jy)&Separation(mas)&$\theta$(deg) &Major(mas) &Axial
ratio &$\phi$(deg) &T$_b$(k)\\
(1)&(2)&(3)&(4)&(5)&(6)&(7)&(8)\\\hline
 \multicolumn{8}{c}{2.3 GHz, $\chi_{\nu}^2=1.02$}\\\hline
 k    &$0.532\pm0.065$ &0      &0        &$0.52\pm0.10$    &1.0 &$0$ &$1.39\times10^{12}$\\
 a    &$1.114\pm0.11$  &$1.55\pm0.16$  &$32\pm3.3$  &$1.04\pm0.11$ &1.0 &$0$ &$7.29\times10^{11}$\\
 f    &$0.318\pm0.034$  &$4.48\pm0.56$  &$25\pm2.6$ &$1.56\pm0.17$ &1.0 &$0$ &$9.25\times10^{10}$\\
 g    &$0.052\pm0.006$  &$8.9\pm0.9$    &$27\pm3.2$  &$3.40\pm0.37$ &1.0 &$0$ &$3.18\times10^{9}$\\
 n1   &$0.022\pm0.004$  &$19.7\pm2.0$   &$29\pm3.3$  &$8.80\pm0.91$ &1.0 &$0$ &$2.01\times10^{8}$\\\hline
    \multicolumn{8}{c}{ 5.0 GHz, $\chi_{\nu}^2=0.78$}\\\hline
 k &$1.908\pm0.19$  &0     &0     &$0.54\pm0.05$     &$0.50\pm0.01$  &$31\pm3.1$ &$1.96\times10^{12}$\\
 a    &$0.347\pm0.036$  &$1.39\pm0.14$  &$41.2\pm4.1$ &$0.92\pm0.09$ &$0.32\pm0.02$ &$6\pm0.6$ &$1.92\times10^{11}$\\
 f    &$0.188\pm0.019$  &$3.65\pm0.63$ &$25.0\pm8.2$ &$2.7\pm0.3$  &$0.57\pm0.01$ &$19\pm2$ &$6.78\times10^{9}$\\
 g    &$0.027\pm0.003$  &$9.18\pm0.93$   &$28\pm2.8$ &$13.79\pm1.5$ &$0.16\pm0.02$ &$34\pm3.4$ &$1.33\times10^{8}$\\\hline
 \multicolumn{8}{c}{ 8.4 GHz, $\chi_{\nu}^2=0.91$}\\\hline
 k &$1.792\pm0.18$  &0     &0     &$0.33\pm0.03$       &$0.43\pm0.01$ &$47\pm4.7$ &$2.03\times10^{12}$\\
 a    &$0.390\pm0.039$  &$1.22\pm0.12$  &$41.4\pm4.1$ &$1.70\pm0.17$ &$0.34\pm0.01$ &$28\pm2.7$ &$2.11\times10^{10}$\\
 f    &$0.069\pm0.007$  &$4.2\pm0.42$ &$24\pm2.4$ &$1.78\pm0.18$ &$0.25\pm0.07$ &$79\pm8.2$ &$4.62\times10^{9}$\\
 g    &$0.018\pm0.003$  &$8.4\pm0.9$   &$22\pm2.2$ &$4.21\pm0.52$  &1.0 &$0$ &$5.39\times10^{7}$\\\hline
 \multicolumn{8}{c}{15.4 GHz, $\chi_{\nu}^2=1.57$}\\\hline
 k &$1.405\pm0.14$ &0 &0 &$0.16\pm0.02$ &1.0 &$0$ &$8.67\times10^{11}$\\
 n2   &$0.521\pm0.052$ &$0.29\pm0.03$ &$45\pm5$ &$0.26\pm0.03$ &1.0 &$0$ &$1.22\times10^{11}$\\
 a    &$0.179\pm0.018$ &$1.42\pm0.14$ &$42\pm4$ &$0.83\pm0.08$ &1.0 &$0$ &$4.10\times10^{9}$\\
 f    &$0.064\pm0.006$ &$3.59\pm0.36$  &$25\pm3$ &$1.75\pm0.18$ &1.0 &$0$ &$3.30\times10^{8}$\\\hline
 \multicolumn{8}{c}{22.2 GHz, $\chi_{\nu}^2=1.83$}\\\hline
 k &$1.156\pm0.17$ &0 &0 &$0.12\pm0.02$ &$1.0$ &$0$ &$6.10\times10^{11}$\\
 n2   &$0.347\pm0.05$ &$0.28\pm0.04$ &$50\pm8$ &$0.23\pm0.03$ &1.0 &$0$ &$4.98\times10^{10}$\\
 a    &$0.104\pm0.016$ &$1.58\pm0.24$ &$39\pm6$ &$0.80\pm0.12$ &1.0 &$0$ &$1.23\times10^{9}$\\\hline

\end{tabular}
\\
Notes: Col. (1): component registration, col. (2): component's flux
density in Jy, col. (3): component's separation from core component
k in mas, col. (4): component's position angle relative to k from
north to east in degree, col. (5): major axis of elliptical Gaussian
component in mas, col. (6): ratio of the minor to major axis of
elliptical Gaussian component, col. (7): position angle of the major
axis of the elliptical Gaussian component from north to east in
degree, col. (8): brightness temperature in K according to the
formula given by Shen et al. (1997).

\end{minipage}
\end{table*}

\section{Results}

The main results are given in this section beginning with the jet
components and an analysis of the jet morphology in Sect. 3.1. Then
in Sect. 3.2, the core is identified through an analysis of the
spectra of the jet components, and we also give the fitting to the
spectrum of the core. In Sect. 3.3, we study the kinematics of three
jet components and discuss the ejections of the jet components. Last
in Sect. 3.4, we try to use the helical model in S95 to describe the
jet structure.

\subsection{Registrations of jet components and morphology}

After carefully analyzing the positions of each component at the
five frequencies the registrations of the components are given in
the first column of Table~\ref{tab:1}. From Fig.~\ref{fig:1} and
Table~\ref{tab:1}, we can see that component k in the south end is
the most compact one and has the highest brightness temperature. B99
assumed it is the core, but we confirm it as the core from its
spectral characteristics in the next subsection. The reason that n1
component is not detected at frequencies higher than 2.3 GHz is
probably due to its extended and weak structure. The separations of
component n2 from k at both 15.4 and 22.2 GHz are less than 0.3 mas.
By considering that it is bigger than 0.3 mas size at frequencies
lower than 15.4 GHz, we think that k component at 2.3, 5.0, and 8.4
GHz in fact contains n2 component. Therefore the flux densities at
2.3, 5.0, and 8.4 GHz for k component actually include n2 emission.
We treat n2 as a new component. Our VLBA maps at 15.4 and 22.2 GHz
also indicate that there seems to be a new but quite weak component
between components a and n2, about 0.65 mas from the core at P.A.
$\sim$ 50$\degr$. However, the inclusion of this component affects
the model fitting little, so we defer this to future observations.
We also note that the map at 5~GHz in Fig.~1 shows a very elongated
component g. We tried to add an additional component in the model
fitting. The fitted new component is very weak ($\sim$5~mJy) and has
little effect on the fitting results. Thus, for simplicity we adopt
component g only to represent the overall extended structure at
5~GHz (see Table 1).

It can be seen from the maps that the overall morphologies at each
frequency are similar, they all extend in the north east direction
with a moving tendency along the P.A.$\approx$$25\degr$. It can also
be seen from Table~\ref{tab:1} that the jet's P.A. changes from
about $50\degr$ at a separation from k component about 0.3 mas to
$40\degr$ at a separation about 1.4 mas and then to $25\degr$ at a
separation about 8 mas, clearly indicating the presence of jet
bending. K95 obtained maps at 8, 22, 43, and 86 GHz, which enabled
them to trace the jet down to sub-mas scale, and found a
60-90$\degr$ misalignment between the sub-mas and mas-structures.
B99 found from their 8-year-long monitoring data of 8 GHz geodetic
VLBI observations that the bending appears more pronounced near the
core area ($\leq$1.5 mas) than farther away, and the jet appears to
bend twice within the inner $\sim$1.5 mas. These results are
consistent with ours, except that our P.A. misalignment of 25$\degr$
is less than that seen by K95. Combined with the data in
Table~\ref{tab:3} in B99, we can see that almost every component's
P.A. decreases with the increasing separation from k until the P.A.
reaches about 20$\degr$. We refer this to as the position angle
regression. In Sect. 3.4 we will try to interpret the position angle
regression using the helical model proposed by Steffen et al.
(\cite{Steffen95}, hereafter S95).

\subsection{Identfication of the core and the spectra of
other components}

\begin{figure*}
\resizebox{\hsize}{!}{\includegraphics{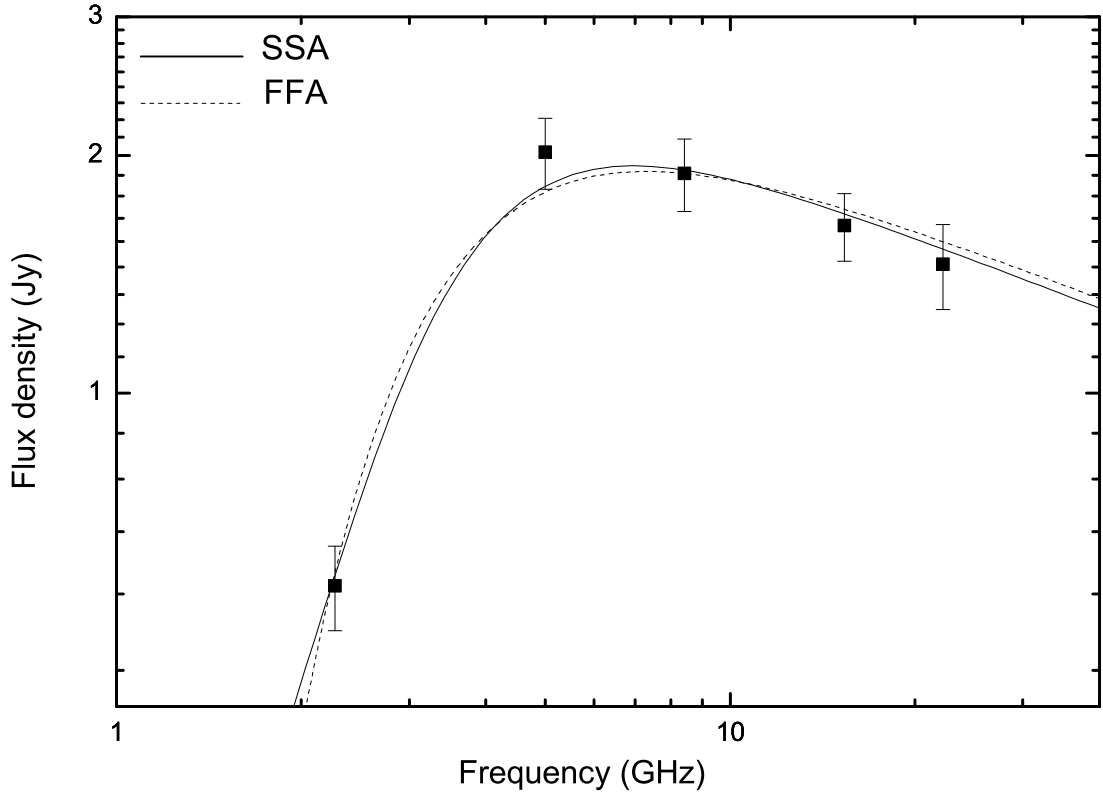}} \caption{The
fitted spectra of component k using the SSA (solid line) and FFA
(dashed line) models.} \label{fig:2}
\end{figure*}

\begin{figure*}
\resizebox{\hsize}{!}{\includegraphics{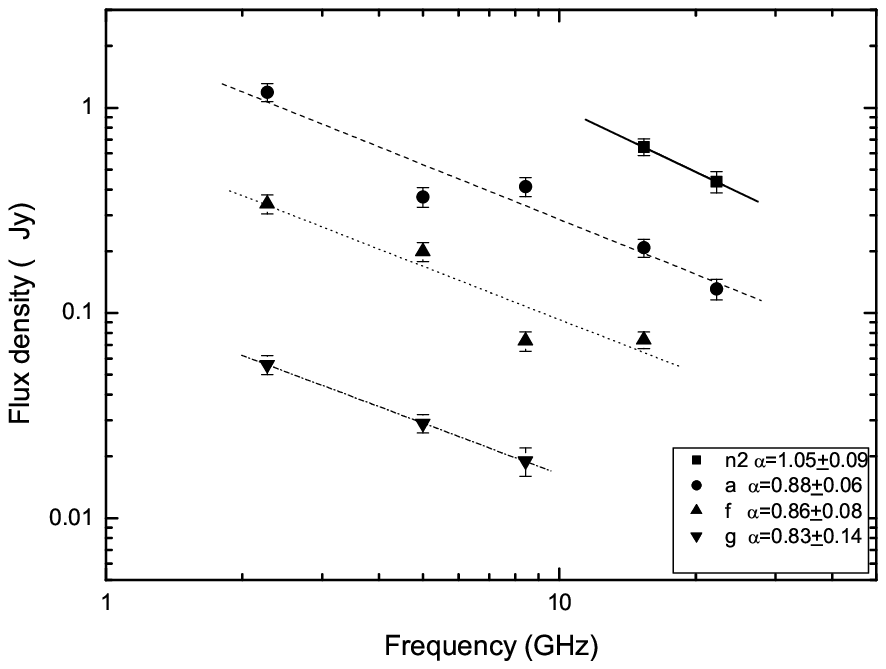}} \caption{The
spectra of VLBI components in \object{PKS 0528+134}. Different lines
represent the linear regressions performed on different VLBI
components, i.e., n2, a, f, and g from top to bottom.} \label{fig:3}
\end{figure*}

The multi-frequency data can be used to study the spectra of the jet
components. The spectra of the k component and other components are
plotted in Fig.~\ref{fig:2} and Fig.~\ref{fig:3}, respectively. Note
that we have applied the absolute flux density correction when
calculating the spectral indexes. There is a spectral turnover for
the component k (Fig.~\ref{fig:2}). Its spectral index between 15.4
and 22.2 GHz is $\alpha=0.30\pm0.08$, much flatter than any other
components (Fig.~\ref{fig:3}). Component k is thus identified with
the core. The identification of the core is important because it is
thought to be related to the central engine in the AGN, and in
general the jet is thought to be ejected from the core, powered by
the central engine that is responsible for the activity of AGN.
Other VLBI components (n2, a, f, and g) all have a typical,
optically thin spectra (Fig.~\ref{fig:3}), consistent with the core
identification. The flux density of n2 at 2.3, 5.0, and 8.4 GHz,
extrapolated from its spectral measurements at 15.4 and 22.2 GHz
without considering the spectral turnover, would be 4.76, 2.10, and
1.22 Jy, respectively. At 2.3 GHz, this is much greater than the
flux density of 0.53 Jy for component k, which is contradicted by
the emission of n2 being included in the emission of k at 2.3 GHz.
Thus, the spectrum of n2 must be inverted at a frequency higher than
2.3 GHz.

Furthermore, we simply try to fit the spectrum of the k component by
using the synchrotron self-absorption (SSA) and free-free absorption
(FFA) formula. According to Kameno et al. (\cite{Kameno00}), the SSA
formula is
$$S_{\nu}=S_0\nu^{2.5}[1-exp(-\tau_s\nu^{-\alpha-2.5)}],\eqno(1)$$
where $S_0$ is the flux density in Jy at 1 GHz when the SSA optical
depth $\tau_s$ at 1 GHz $\gg1$, and $\alpha$ is the optically thin
spectral index. The FFA formula is
$$S_{\nu}=S_0\nu^{-\alpha}exp(-\tau_f\nu^{-2.1}),\eqno(2)$$
where $S_0$ is the flux density in Jy at 1 GHz extrapolated from the
optically thin spectrum, and $\tau_f$ is FFA optical depth at 1 GHz.
Assuming $\alpha$ is known to be 0.30, we list the fitting results
in Table~\ref{tab:2} with the corresponding fitting curves also
plotted in Fig.~\ref{fig:2}. The inverted spectrum can be fitted
equally well by both absorption models. It can be seen in
Fig.~\ref{fig:2} that the spectral turnover happens at about 6.95
GHz with the corresponding flux density 1.93 Jy. Using the relation
$\frac{\nu_m}{1~ GHz}\sim8(\frac{B}{1~
G})^{\frac{1}{5}}(\frac{S_m}{1~Jy})^{\frac{2}{5}}(\frac{\theta}{1~mas})^{-\frac{4}{5}}(1+z)^{\frac{1}{5}}$
for the SSA model (Kellermann \& Pauliny \cite{Kellermann81}), we
can estimate a constant magnetic field $B=0.35$ mG in the core
region. Here $\nu_m=6.95$ GHz, $S_m=1.93$ Jy and $\theta=0.30$ mas
are used. Following Marscher (\cite{Marscher87}), we adopt
$\theta=\sqrt{\theta_a \theta_b}$ as the angular diameter of the
core, where $\theta_a$ and $\theta_b$ are the major and minor axes
of the Gaussian component. Since there is no VLBI size measurement
at $\nu_m=6.95$ GHz, the adopted $\theta=0.30$~mas is an arithmetic
average of the sizes at 5.0 and 8.4 GHz. The synchrotron cooling age
is estimated to be at least $2.9\times10^4$~yrs using the expression
$t_{1/2}=2.76\times 10^4~
(yr)~(\frac{B}{1~mG})^{-1.5}(\frac{\nu^{\prime}_m}{1~GHz})^{-0.5}\sin^{-1.5}\alpha$
(e.g., Rybicki \& Lightman \cite{RL79}). Here, B=0.35~mG, the
turnover frequency in the source rest frame
$\nu_m^{\prime}$=21.33~GHz (after correcting the cosmological
redshift effect), and the angle between the electron velocity and
magnetic filed $\alpha$=$\frac{\pi}{2}$.

\begin{table*}
\centering \caption{The fitting parameters of Fig. 2.} \label{tab:2}
\vspace{2mm}
\begin{tabular}{cccc}\hline\hline

Model   &$S_0$(Jy)    &$\tau$   &$\chi^2$\\\hline
SSA  &$0.076\pm0.006$  &$51\pm5$  &1.17 \\
FFA  &$4.0\pm0.2$  &$9.3\pm0.6$  &1.78\\\hline

\end{tabular}
\end{table*}

\subsection{Proper motion and component ejection}

\begin{figure*}
\centering \resizebox{\hsize}{!}{\includegraphics{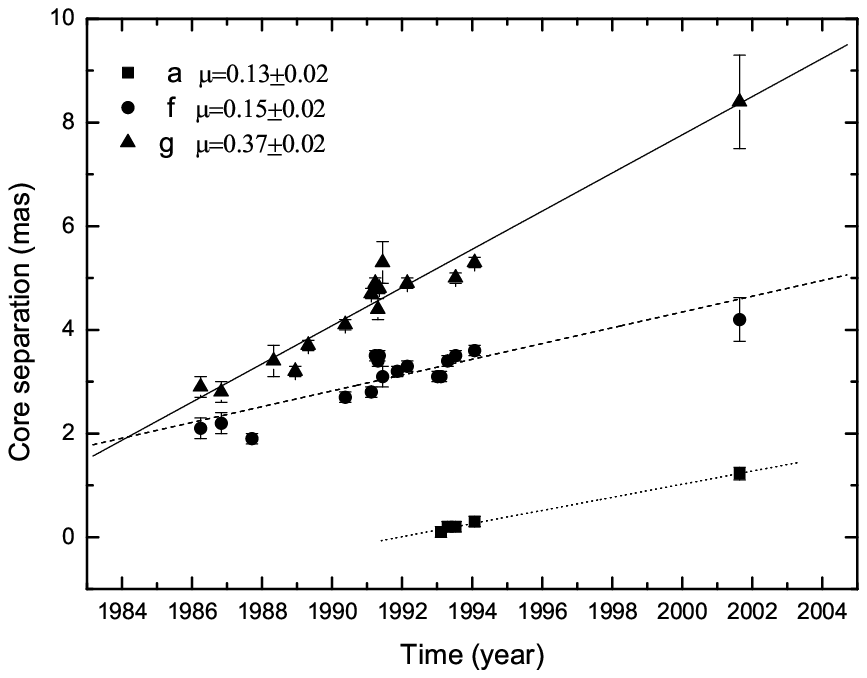}}
\caption{The linear fitting to the separations of components a, f,
and g from the core (component k). The data before 2001.64 are from
B99, the ones at 2001.64 are ours at 8.4 GHz.} \label{fig:4}
\end{figure*}

To find the relations between the seven components in B99 and the
components in our paper, we extrapolate the proper motions of the
seven components in B99 to our observing time and find that the
positions of a, f, and g components in B99 are near the positions of
our a, f and g, so we think that our a, f, and g are the same as a,
f, and g in B99. This is also the reason we used the same component
registration. The counterparts to other components (b, c, d, and e)
in B99 are not found in our VLBA observations. Figure~\ref{fig:4}
shows the linear fit to the separations from the core of these VLBI
jet components a, f, and g at 8.4 GHz. All the data before 2001.64
are from B99. Table~\ref{tab:3} lists the proper motion values of
each component; due to the non-detections of b, c, d, and e in our
observations, the proper motion values of components b, c, d, and e
are from B99, but their apparent velocities are re-calculated using
the new cosmological constants ($H_0=71$ kms$^{-1}$
 Mpc$^{-1}$, $\Omega_M=0.27$, and
$\Omega_\Lambda=0.73$). It can be seen from Table~\ref{tab:3} that
the proper motion of the outmost component g (see Fig.~\ref{fig:1})
becomes about 3 times faster than those of other inner jet
components a, b, c, d, e, and f, which are essentially the same
about 10.2-12.7c within errors.

\begin{table*}
\centering \caption{Combined with the data from B99, the proper
motions are calculated assuming linear outward motion of the
components. } \label{tab:3} \vspace{2mm}
\begin{tabular}{ccccccccc}\hline\hline

ID. &a  &b  &c  &d  &e  &f  &g\\\hline $\mu({\rm mas/yr})$
&$0.13\pm0.02$ &$0.14\pm0.03$ &$0.12\pm0.01$ &$0.13\pm0.01$
&$0.13\pm0.01$ &$0.15\pm0.02$
&$0.37\pm0.02$\\
$\beta_{app}$ &$11.0\pm1.7$ &$11.8\pm2.5$ &$10.2\pm0.8$
&$11.0\pm0.8$ &$11.0\pm0.8$ &$12.7\pm1.7$ &$31.3\pm1.7$\\\hline

\end{tabular}
\end{table*}

Usually, the ejection of a new VLBI component can be related to a
flux outburst in the light curve (e.g. K95; B99), therefore, we also
relate the ejection of n2 to one burst. Assuming that there is no
big difference in proper motion between n2 and other components, we
adopt 0.13 mas/yr (see Table~\ref{tab:3}) as the proper motion for
n2. Thus the ejection (zero-separation) epoch of n2 is roughly
around 1999.5. This just falls into a burst duration from 1999 to
2000.5 shown on the well-sampled light curves of \object{PKS
0528+134} at 22 and 37 GHz (Ter\"{a}sranta et al.
\cite{Terasranta05}), suggesting that the ejection of n2 may be
related to this radio burst, which peaks at about 2000. Likewise,
the ejection epoch of component a is estimated to be about
1991.94$\pm$0.18 from its proper motion 0.13 mas/yr. This ejection
epoch is consistent with the 1992.1$\pm$0.4 found by B99. Based on
the radio light curve at 8 GHz (Fig.~\ref{fig:5} in B99), B99 found
that the ejection epoch of component a was near the beginning of the
radio burst peaking around 1993.6, so they related the ejection of
component a to this radio burst, which is preceded by a gamma-ray
burst peaking at about 1993.23 (Mukherjee et al.
\cite{Mukherjee99}). But the light curves at 22 and 37 GHz
(Ter\"{a}sranta et al. \cite{Terasranta05}) show another radio burst
peaking at about 1992.32, and the flux-density variation at
millimeter-bands shows a flux-density burst at about 1991.94 (K95),
so we cannot exclude the possibility that the radio burst peaking at
about 1992.32 or 1991.94 may be related to the ejection of component
a. The relations between the ejection of a new VLBI component and
the flux-density bursts at radio and gamma-ray bands deserve more
study to ascertain which flux-density burst is related to the
ejection of the new VLBI component.

\subsection{Helical pattern of the jet structure}

VLBI observations have revealed a helical structures in some AGNs,
e.g., 3C 345 (S95), 3C 120 (Hardee et al. \cite{Hardee05}), and Mkn
501 (Villata \& Raiteri \cite{Villata99}). Two bendings of the jet
in \object{PKS 0528+134} on the mas scale indicate the possible
presence of the helical trajectory in \object{PKS 0528+134}. The
helical structure may result from the orbital motion in a binary
black-hole system, Newtonian precession, internal jet rotation, or
global helical magnetic fields (e.g. Rieger \cite{Rieger05}). S95
discussed the jet component's kinetics in detail through their
helical models by using the conservation laws for kinetic energy,
angular momentum, momentum, and jet opening angle.

Any three conservation quantities can give a jet trajectory, so
there are four possible models. The first model assumes the
conservations of momentum along the jet axis $p_z$, the kinetic
energy $E_{kin}$, and the specific angular momentum $L_z$, and gives
a rotational hyperboloid with less than a quarter revolution. The
second model assumes the conservations of $p_z$, $E_{kin}$, and the
opening angle $\psi$, and gives a helical trajectory; but the helix
is self-similar in space and time, and its helical amplitude is not
dampened. The third model assumes the conservations of $E_{kin}$,
$L_z$, and $\psi$, and gives a helical trajectory that is not
self-similar but has an asymptotic behavior; i.e., the pitch angle
of the helix is the smallest at the origin, and it opens along the
jet axis. The fourth model assumes the conservations of $p_z$,
$L_z$, and $\psi$ and gives a similar trajectory to the one in the
third model. The major difference between the third and fourth
models lies in how the third model predicts a high level of
component speeds as a component separates from the core. S95 applied
the third model to fit the trajectories of C4 and C5 in 3C 345
simply because this model reflects the basic character of the
trajectories and motion observed in 3C 345. From Sect. 3.1 we know
that the jet in \object{PKS 0528+134} experiences two bends, and
there is a position-angle regression phenomenon for the P.A.s of the
jet components, so obviously the first and second models are not
applicable to the case of \object{PKS 0528+134}. From
Table~\ref{tab:3} we can find that the apparent velocities of the
jet components of \object{PKS 0528+134} remain on a high level as
the components separate from the core and appear to increase with
distance from the core, so we reject the fourth model and adopt the
third model to describe the bending trajectory of the jet in
\object{PKS 0528+134}.

\begin{figure*}
\centering
\includegraphics[width=12cm]{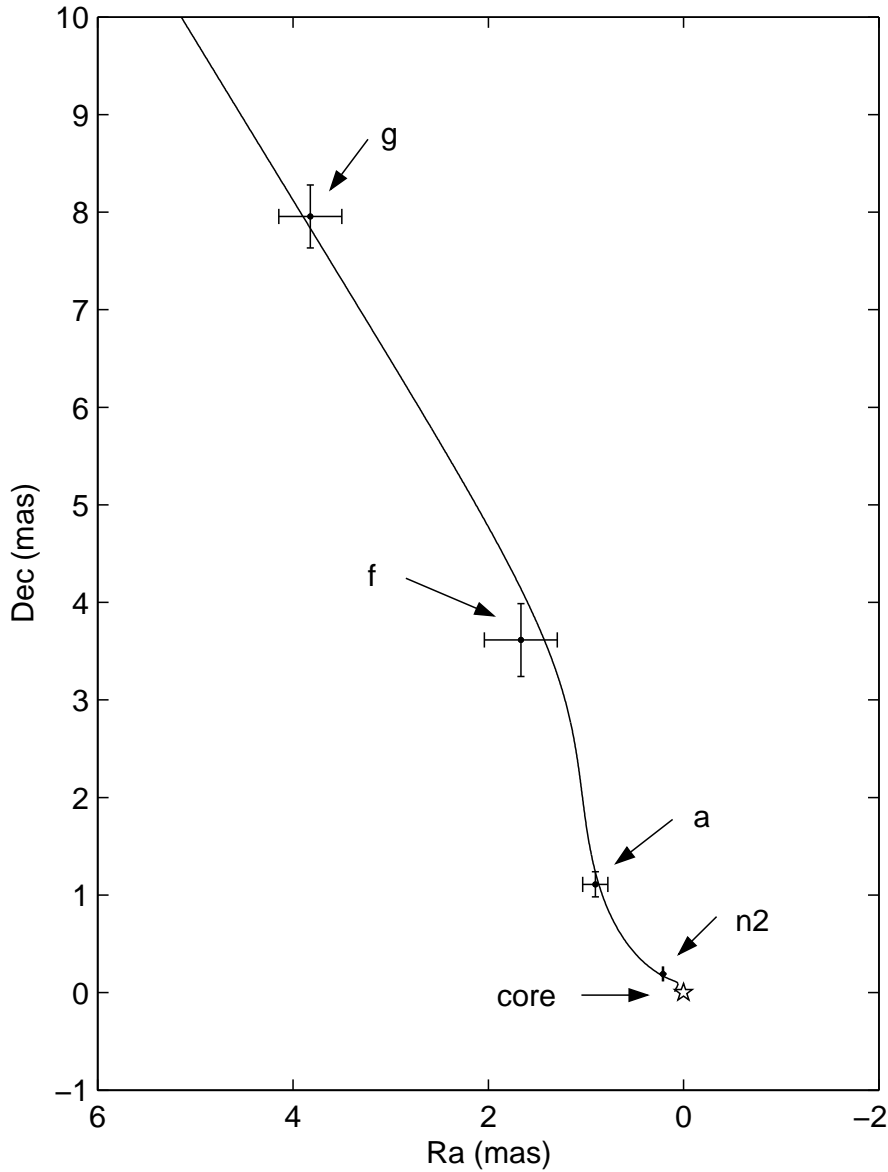}
\caption{We use the third model of Steffen et al.'s (1995) helical
models to approximately fit the trajectory of the jet components
obtained from our quasi-simultaneous five-frequency VLBA maps. The
locations of the components relative to the core (component k in
Fig. 1) are the average values of the locations measured at
different frequencies. The standard deviation of the locations of
each component relative to the average is adopted as the error bar.}
\label{fig:5}
\end{figure*}

The corresponding sets of equations for the third model in the
cylindrical coordinates $r,z$, and $\phi$, are:
\setcounter{equation}{2}
\begin{eqnarray}
r(t)&=&\sqrt{(at+b)^2+c^2}~,  \\
z(t)&=&\frac{r(t)-r_0}{tan \psi}~,\\
\phi(t)&=&\phi_0+\frac{1}{sin
\psi}(arctan\frac{at+b}{c}-arctan\frac{b}{c})~,
\end{eqnarray}
where $a=\beta$sin$\psi$, $b=\sqrt{r_0^2-L_z^2/\beta^2}$,
$c=L_z/\beta$, $\beta$ is the jet component's intrinsic velocity in
units of the light speed, and $r_0,z_0$, and $\phi_0$ are the
initial coordinates of the jet component (for simplicity, we set
$z_0=0$). The best fit is not given here because the currently
available observational data points are not enough to constrain the
evolution of the VLBI jet position with time, which requires 10
parameters to describe (S95). We show an approximate fit in
Fig.~\ref{fig:5}, which is obtained through gradually adjusting the
fitting parameters to minimize the difference between the fitted
trajectory and the observed data points. Some important fitting
parameters are: the angle between the line of sight and the axis of
cone is 3$\degr$, the intrinsic velocity $\beta=0.9995$, the conical
half opening angle $\psi=0.25\degr$, $r_0=0.11$ mas, and the initial
angular velocity $\omega_0=-0.80$$\degr$/yr. According to the
relativistic beaming model, $\beta_{app}=\frac{\beta
sin\theta}{1-\beta cos\theta}$, where $\beta_{app}$ is the apparent
velocity in units of the speed of light, and $\theta$ is viewing
angle to the moving direction of the jet. The allowable maximal
value of $\theta$ is $\theta_{max}=2\arctan(\frac{1}{\beta_{app}})$
(Pearson \& Zensus \cite{Pearson87}), so when $\beta_{app}$=13,
$\theta_{max}$=8.8$\degr$. Therefore, our choice of a viewing angle
of 3$\degr$ in Fig.~\ref{fig:5} is reasonable. We can also estimate
a Doppler factor from $\delta\sim\frac{1}{sin\theta}$. When
$\theta=3\degr$, this gives $\delta\sim19$, close to the value of 17
from $\delta=\frac{1}{\Gamma(1-\beta{cos\theta})}$
($\Gamma=\sqrt{\frac{1}{1-\beta^2}}$ is Lorentz factor) when using
 $\beta=0.9995$. We thus feel that the fit shown in Fig.~\ref{fig:5} is physically
 reasonable.

Furthermore, this model can be used to explain the position-angle
regression discussed in Sect. 3.1. Since $L_z=r\beta_{\omega}$ is
conservative ($\beta_{\omega}$ is the velocity of the jet component
rotating around the conical axis), $\beta_{\omega}$ is decreasing
with the increasing r when the jet component is moving out along the
conical surface. By differentiating Eq. (3), one can obtain the
radial velocity component
$\beta_r=\frac{a(at+b)}{\sqrt{(at+b)^2+c^2}}$, which is
perpendicular to $\beta_z$ and $\beta_{\omega}$ and which increases
with the time. Similarly, from Eq. (4), one can get
$\beta_z=\frac{\beta_r}{tan\psi}$, which increases much faster than
$\beta_r$ as tan$\psi\ll1$; i.e., $\beta_{\omega}$ is mainly
converted to $\beta_z$. There is thus a reacceleration process of
the velocity component $\beta_z$ moving along the jet axis, and
finally the motion of the jet component is dominated by $\beta_z$.
Thus, the jet component moves just about along the conical axis, and
the P.A. barely varies when the jet component moves into the outer
region of the jet; i.e., the moving direction on the map of the jet
component regresses to the projected direction of the cone axis on
the sky.

\section{Discussion}

Our quasi-simultaneous VLBI observations at five frequencies allow
us to study the spectral distribution of \object{PKS 0528+134} for
the first time. From the light curves at 22 and 37 GHz
(Ter\"{a}sranta et al. \cite{Terasranta05}), we can see that our
observing time 2001.64 \object{PKS 0528+134} is in the quiescent
state, so the spectral distribution represents the spectral
distribution of \object{PKS 0528+134} in its quiescent state. The
identification of the core in VLBI analysis is important because the
registration, tracing, and kinematic analysis of the jet components
are all based on such a core identification. B99 assumed that the
brightest and most compact component at the southern end of the jet
was the core, based on how all other VLBI components seemed to move
away from it, and how it showed a pronounced flux density variation
and an inverted spectrum (Pohl et al. 1995).

Our spectral analysis indicates that component k has a flat spectral
index, $\alpha=0.3\pm0.08$, with the spectrum inverted at 6.95 GHz,
consistent with the k component being the compact core. According to
the SSA theory, the separation of an optically-thin jet component to
the core increases with the increasing frequency (Lobanov
\cite{Lobanov98}). However, due to large uncertainties in the
position determination in our observations, we cannot tell whether
there is such a core-position offset in \object{PKS 0528+134}.
Phase-referenced VLBI observations can achieve a high positional
precision of tens of microarcsecond (Ros \cite{Ros03}) and can thus
be used to study the absorption mechanism in the core region of
\object{PKS 0528+134} through investigating the frequency-dependent
position offset.

By tracing the components in B99, we find our a, f, and g components
are identical to the a, f, and g components in B99. All three
components show high apparent velocities, especially g up to 31.3c.
The ejection times of component a and the newly detected n2
component can be estimated from the extrapolation of the proper
motions to their zero separations from the core. The ejection of n2
at about 1999.5 could be related to the radio burst peaking at about
2000, but we cannot be sure which flux-density burst is related to
the ejection of component a. We can see from the light curves at 22
and 37 GHz given by Ter\"{a}sranta et al. (\cite{Terasranta05}) that
there are six significant flux-density outbursts before our
observation at epoch 2001.64, but only two outbursts could be
related to the ejections of VLBI components. The reason that not
every flux density outburst corresponds to an ejection of a VLBI
component could be that the related VLBI components are too weak or
have reached its radiation loss time before our observations. Also
it is not so clear whether or not every outburst can be related to
the ejection of a VLBI component.

According to the morphology of \object{PKS 0528+134}, we assume for
the first time that it has a helical structure. New VLBA
observations at 15 GHz (Lister \& Homan \cite{Lister05}) have also
confirmed the presence of a bending jet structure in \object{PKS
0528+134}. We used the third helical model in S95 to describe the
jet trajectory in \object{PKS 0528+134}. S95 fitted this helical
model to the trajectories of C4 and C5 in 3C 345. In 3C 345 there is
a component, 18 mas from the core, whose P.A. is different from the
P.A.s of its inner components by 20$\degr$ (Unwin \& Wehrle
\cite{Unwin92}), which is contrary to the general prediction of this
helical model that the jet component will move close to the jet axis
little by little; i.e., there is a position-angle regression.
Because of this, S95 only restricted their analysis to the inner 3
mas from the core. Our VLBA observations at 2.3 GHz
(Fig.~\ref{fig:1}) show that there is no obvious P.A. change for the
external jet components of \object{PKS 0528+134}. We then can apply
this helical model to the jet of \object{PKS 0528+134} until 20 mas
from the core since there is no abrupt P.A. change for the external
jet components as seen in 3C 345. Some core-dominated radio sources
even show 90$\degr$ misalignment between the P.A.s of the VLBI and
VLA scale jets (Conway \& Murphy \cite{Conway93}), which is
difficult to explain by the helical model we adopted. Conway \&
Murphy (\cite{Conway93}) used their low-pitch helical model to
explain this phenomenon, but this model is difficult to explain the
several bends at the VLBI scale in radio sources like \object{PKS
0528+134} and 3C 345. We think that the tremendous change of the
P.A. on a large scale may be due to the interactions between the jet
and ambient medium of the host galaxy or another external force
performed on the jet, so the angular momentum conservation required
by the helical model we used in Sect. 3.4 is violated. Additional
dynamical analysis would be required in order to study the
large-scale jet structure. Except that many extragalactic radio
sources have the helical jets, recent VLBI observations show that
some microquasars in our Galaxy also have these helical patterns.
For example, both GRO J1655-40 (Hjellming \& Rupen
\cite{Hjellming95}) and SS433 (Blundell \& Bowler \cite{Blundell04})
exhibit the helical jets.

We know from the analysis of the helical model that the velocity of
jet component is dominated by the velocity component along the cone
axis, i.e., $\beta_z$. According to the helical jet model, $\beta_z$
should first increase rapidly and then approach its theoretical
value gradually if the angle between the line of sight and the
z-axis is constant. From Fig.~\ref{fig:4} we find it impossible that
the apparent velocities of components a and f are reaccelerated to
that of component g by this helical model. A more realistic
situation could be that the velocity component $\beta_z$s of these
jet components have been accelerated shortly after their emergence,
and that the initial ejected velocity of g component is bigger than
others'. Obviously, this scenario requires a very good time coverage
of high-resolution VLBI observations to test it, especially during
the period when the jet component is initially ejected.

\section{Conclusions}

The five high-resolution VLBI maps made at frequencies 2.3, 5.0,
8.4, 15.4, and 22.2 GHz show that \object{PKS 0528+134} has a jet
structure extending in the north east direction, and the jet bending
is clearly seen at about 0.3 and 1.4 mas from the core. Our
registration of VLBI components a, f, and g is consistent with the
one in B99. We also detect a new VLBI jet component n2, which is
likely to be ejected during the radio burst at about 1999.5. We
estimate the ejection epoch of component a to be 1991.94 by the
linear back-extrapolation of its proper motion, but cannot determine
which flux-density burst is truly related to the ejection of
component a. No counter-jet is seen in our maps, but its existence
cannot be ruled out. We also find that the position angles of the
jet components in \object{PKS 0528+134} are large when they are
initially ejected from the core, but gradually decrease with the
increasing separation from the core until about 25$\degr$, which is
referred to as the position angle regression.

This first quasi-simultaneous VLBA observations at five frequencies
provide a good condition to calculate the spectra of the jet
components without taking the variations of the flux densities with
the time into account. We identify k component as the core mainly
based on its flat spectral index of $0.30\pm0.08$ and an inverted
spectrum peaking at about 6.95 GHz. But we cannot determine which
absorption mechanism (SSA, FFA, or others) is mainly responsible for
the observed inverted spectrum; future low-frequency observation
with sufficient spatial resolution may be useful for distinguishing
these possible models. The brightness temperature $T_b$ of each
component is given in Table~\ref{tab:1}; the brightness temperature
of core component k does not exceed the equipartition brightness
temperature cutoff of $\sim10^{11}$ K (Readhead \cite{Readhead94})
if we take the Doppler boosting effect into account. The presence of
fast superluminal motions indicates that the relativistic effects
are important in this source.

We attempt to describe the trajectory of the jet in \object{PKS
0528+134} using the helical jet model proposed by S95. The
parameters of kinetics obtain from this helical model are consistent
with ones from observations. We find from this helical model that
there is a reacceleration process of the velocity component
$\beta_z$ along the jet axis and that the jet finally moves nearly
along the jet axis, which can be used to explain the position-angle
regression. Since the reacceleration process in this helical model
cannot explain this big difference in the jet's velocities, we
suggest further that a much higher apparent velocity of g component
than its inner components' may be due to the relatively higher
initial velocity of g component than its inner components. To
understand the reacceleraton process well in this helical model
needs a good time coverage of high-resolution VLBI observation just
as the jet component ejects.

\begin{acknowledgements}
We thank the anonymous referee for critical comments and helpful
suggestions on the manuscript. This research made use of data from
the University of Michigan Radio Astronomy Observatory, which is
supported by the University of Michigan and the National Science
Foundation. The Green Bank Interferometer is a facility of the
National Science foundation operated by NRAO with support from the
NASA High Energy Astrophysics program. This work was supported in
part by the National Natural Science Foundation of China (grant
10573029). Z.-Q. Shen acknowledges support by the One-Hundred-Talent
Program of the Chinese Academy of Sciences.
\end{acknowledgements}

\clearpage
\end{document}